\documentclass[11pt]{article}
\usepackage[utf8]{inputenc}
\usepackage[T1]{fontenc}
\usepackage{amsmath}
\usepackage{amsfonts}
\usepackage{amssymb}
\usepackage[version=4]{mhchem}
\usepackage{float}
\usepackage{url}
\usepackage{multirow}
\usepackage{stmaryrd}
\usepackage{graphicx} 
\usepackage{booktabs} 
\usepackage[a4paper, margin=1in]{geometry}

\usepackage[style=apa, backend=biber]{biblatex}
\DeclareLanguageMapping{american}{american-apa}

\graphicspath{ {./Images/} }
\usepackage[hidelinks]{hyperref}
\title{Wavelet Based Cross Correlations with Applications}

\author{Jack Kissell  \and Vijini Lakmini\thanks{Corresponding Author (\tt vijinil@tamu.edu) }  \and Brani Vidakovic }

\date{
	Department of Statistics, Texas A\&M University, College Station, TX, USA
}

\begin{document}

\maketitle
\begin{abstract}
Wavelet Transforms are a widely used technique for decomposing a signal into coefficient vectors that correspond to distinct frequency/scale bands while retaining time localization. This property enables an adaptive analysis of signals at different scales, capturing both temporal and spectral patterns. By examining how correlations between two signals vary across these scales, we obtain a more nuanced understanding of their relationship than what is possible from a single global correlation measure. In this work, we expand on the theory of wavelet-based correlations already used in the literature and elaborate on wavelet correlograms, partial wavelet correlations, and additive wavelet correlations using the Pearson and Kendall definitions. We use both Orthogonal and Non-decimated discrete Wavelet Transforms, and assess the robustness of these correlations under different wavelet bases.  Simulation studies are conducted to illustrate these methods, and we conclude with applications to real-world datasets.
\end{abstract}

\section{Introduction}

Many modern datasets arising in climate monitoring, finance, biomedical engineering, and industrial process control are non-stationary, exhibiting statistical properties that vary over time and frequency. Classical correlation measures, such as the Pearson coefficient, summarize linear dependence between two entire signals under the assumption that this dependence is constant. While analytically convenient, this assumption can obscure relevant structures, particularly when dependencies arise intermittently, are confined to specific frequency bands, or vary across temporal scales.

Wavelet-based cross-correlation provides a framework to address these limitations by integrating multiresolution wavelet analysis with correlation measures. This approach enables characterization of evolving dependence structures simultaneously across time and frequency. Unlike Fourier-based methods, which emphasize frequency resolution at the expense of temporal localization, wavelet transforms retain both. Each wavelet coefficient is associated with a specific scale and time, permitting correlation analysis that is both scale-resolved and time-localized. Consequently, wavelet-based correlation methods are well suited for identifying dependencies that shift in lag, appear selectively within frequency bands, or differ across scales.

Initial developments of wavelet-domain correlation emerged in the mid-1990s. \textcite{HudginsFrieheMayer1993} were first to define cross-scalograms and scale-dependent correlation using the Continuous Wavelet Transforms.  Percival and collaborators at NOAA established related concepts of wavelet variance and covariance 
\parencite{percival1995, percivalWalden2000_wmtsa}. These early contributions emphasized the use of the discrete wavelet transform (DWT) and the maximal overlap discrete wavelet transform (MODWT) to decompose variance of geophysical time series and to construct confidence intervals for scale-dependent estimates. \textcite{Lindsayetal1996} further developed scale-dependent measures of variance and covariance, applying both the DWT and MODWT with Daubechies (LA8) filters to sea surface temperature data from the Beaufort Sea. Subsequently, \textcite{Whitcher2000} extended this line of work to wavelet cross-covariance and cross-correlation, identifying correlations and lead-lag relationships in bivariate atmospheric series at specific temporal scales while controlling for additional seasonal factors.

Subsequent methodological and applied studies broadened the scope of wavelet correlation. \textcite{arsenievMalykhinaKratirov2024} employed wavelet cross-correlation in a dual $\gamma$-ray meter for oil well monitoring, isolating scale-specific peaks associated with liquid and gas velocities even under low signal-to-noise conditions. In astrophysics, \textcite{frick2001scaling} proposed a two-dimensional extension using a Mexican Hat wavelet basis to quantify correlations between galactic images at different wavelengths, revealing structural relationships undetectable by pixel-level correlation. In statistical methodology, \textcite{souzaFelix2018_tema} studied bivariate time series via wavelet cross-correlation using the non-decimated wavelet transform (NDWT). Their analysis demonstrated that traditional cross-correlation can yield biased lag estimates for autocorrelated or non-stationary signals, whereas wavelet-based approaches provided sharper confidence intervals and improved identification of scale-specific lags. In economics and finance, \textcite{fernandezMacho2012_physicaA} introduced Wavelet Multiple Correlation and Wavelet Multiple Cross-Correlation, extending analysis to multivariate systems. Applied to Eurozone stock market data, these methods revealed strong long-scale correlations across markets and suggested leading–lagging structures among variables. \textcite{Gallegati2008WaveletAnalysis, Gallegati2012WaveletContagion} applied wavelet correlations in the context of analysis of microeconomic time series.
\textcite{Schnaidt2019PhD} also provides some theoretical properties of wavelet correlations.

These developments illustrate the methodological richness and practical utility of wavelet-based cross-correlation. By enabling time- and scale-resolved analysis of dependence, wavelet methods generalize classical correlation and uncover structures that remain hidden under stationarity assumptions. The resulting framework has proven useful across diverse fields where evolving, scale-specific dependencies are central to understanding complex dynamical systems.

\section{Methodology}
Our methodology covers both orthogonal and non-decimated wavelet transforms in one and two dimensions. In addition to wavelet correlation, we define wavelet partial correlation and wavelet correlograms. We also investigate  Kendall’s $\tau$, a robust correlation measure, which is sensitive to  monotonicity that may not be linear.

\subsection{Orthogonal and Nondecimated Wavelet Transforms via Matrices}

The simplest way to introduce wavelet transforms is via \emph{linear transforms} defined by so-called \emph{wavelet matrices}. This perspective emphasizes that a wavelet transform is a change of basis in $\mathbb{R}^n$, similar in spirit to the Fourier transform but constructed from localized basis functions.

For a \emph{discrete orthogonal wavelet transform (DWT)}, the wavelet matrix $W$ is constructed directly from the \emph{low-pass} and \emph{high-pass} filter coefficients associated with the chosen wavelet family. The orthogonality of the filters ensures that the resulting matrix is orthogonal as well. If the input data are arranged as a vector $y \in \mathbb{R}^n,$  the forward and inverse wavelet transforms are expressed as
\begin{equation}
\label{eq:worth}
d = Wy, \qquad y = W^\top d,
\end{equation}
where  $d$ is the vector of wavelet coefficients (containing both scaling and detail coefficients), $y$ is the data vector, and $W$ is the orthogonal wavelet matrix satisfying $W W^\top = W^\top W = I_n$. To be precise, for a vector $y$ of dyadic length $n=2^J$, the transform of the depth $L$ results in a vector $d$ of the form
\begin{eqnarray}
    \label{eq:form}
    d=\left(c_{J-L}, d_{J-L}, d_{J-L+1}, \dots, d_{J-1}\right),
\end{eqnarray}
with indexes representing multiscale levels.
Here, $c_{J-L}$ is a subvector of
scaling coefficients (a smooth approximation) and $d_{J-i}$ are vectors of detail coefficients, with $d_{J-1}$ being the vector of finest details, and $d_{J-L}$  the vector of coarsest details. 
The length of a subvector with index $J-i$ is $2^{J-i},$ so the length of $d$ being $2^{J-L}+\sum_{i=1}^{L} 2^{J-i} = 2^J= n,$ as expected.

Thus, the transformation is perfectly reversible: no information is lost in going back and forth between the data and its wavelet representation. The structure of $W$ depends on the number of decomposition levels chosen. The resulting $W$ is always an $n \times n$ matrix for data of length $n$. Simple functions such as \texttt{Wavmat.m} (MATLAB) or \texttt{Wavmat.py} (Python), available in the repository associated with 
\textcite[p.~115-117]{Vidakovic1999}
describes construction of $W$, step-by-step.

This matrix-based approach is operationally convenient and intuitive, since one can view the transform as a single linear operator. However, the computational cost of explicitly forming and multiplying by $W$ grows as $\mathcal{O}(n^2)$, which can be prohibitive for large $n$. For such cases, the fast wavelet transform (Mallat’s pyramid algorithm, \textcite{Mallat1989}) is used instead, reducing the complexity to $\mathcal{O}(n)$.

The same matrix-based perspective extends naturally to the  nondecimated wavelet transform (NDWT), also known as the stationary wavelet transform. Unlike the orthogonal transform, the NDWT does not downsample after filtering. Instead, it inserts zeros (upsampling) into the filters at each level, ensuring that the levelwise coefficient sequences remain of length $n$.

If $L$ levels of decomposition are performed on a signal of length $n$, the resulting transform can be represented as
$$ 
d = Wy,
$$
where now $W$ is of size $(L+1)n \times n$. At each level, all coefficients are retained (no decimation). The form of $d$ is the same as in (\ref{eq:form}), but the length of each subvector is $n$, so the length of $d$ is $(L+1) n.$ As in the case of orthoginal transform, the rows of $W$ are built from shifted and dilated versions of the low-pass and high-pass filters, but augmented with zeros according to the level, see \parencite{nason1995stationary}.

While in the NDWT case $W$ is no longer orthogonal, it still defines a structured linear operator with several advantages:
(i) it produces redundant, translation-invariant representations of signals,
(ii) it makes statistical modeling more precise, since the transform preserves the alignment with the original time indices, and
(iii) it facilitates theoretical analysis by providing a unified matrix-based view of the transform.

A detailed description of how to construct NDWT matrices can be found in \textcite{kang2016wavmatnd}, where explicit constructions and examples are provided.
 \vspace{0.2in}
\subsection*{2D Wavelet Transforms}

There are several ways to define the two-dimensional (2D) wavelet transform. 
Among them, the formulation based on wavelet matrices is often the most intuitive. 
Let $A$ be an image of size $m \times n$. 
Then the orthogonal 2D wavelet transform of $A$ can be written as
\begin{eqnarray*}
    D = W_1 A W_2^\top,
\end{eqnarray*}
where $W_1$ and $W_2$ are orthogonal wavelet matrices of size $m \times m$ and $n \times n$, respectively, and the transform $D$ is a matrix of size $m \times n$. 
This transform is sometimes referred to as a \emph{scale-mixing transform}, since the tessellation of $D$ contains rectangular blocks with coefficients coming from different scales along the $x$ and $y$ axes.

In most applications, the input images are square of size $n \times n$, so that $W_1$, $W_2$, and $D$ are all $n \times n$ matrices. 
The tessellation of $D$ contains a diagonal hierarchy where scales coincide in both directions. 
This diagonal hierarchy corresponds to the standard 2D wavelet transform, where the decomposition is implemented symmetrically in both directions,
\parencite[p. 256]{Daubechies1992}. 

In this paper, we focus not on wavelet correlations in the orthogonal 2D setting, but rather on those defined in the nondecimated framework. 
Analogous to the orthogonal case, the 2D nondecimated wavelet transform of an image $A$ of size $n \times n$ is defined as
\begin{eqnarray*}
    D = W A W^\top,
\end{eqnarray*}
where $W$ is a nondecimated wavelet matrix of size $(L+1)n \times n$, and $D$ is the corresponding matrix of wavelet coefficients of size $(L+1)n \times (L+1)n$. 

The matrix $D$ contains $(L+1)$ diagonal submatrices of which $L$ blocks correspond to detail coefficients at different scales, and one block corresponds to the scaling (or smooth) coefficients. 
On this diagonal hierarchy, we define wavelet correlations between two input images (or matrices) $A$ and $B$ of the same size. 
As in the univariate case, one may employ different definitions of correlation, extend to partial correlations, and construct associated confidence intervals. These definitions and extensions will be discussed next.

\vspace*{0.2in}

\subsection*{Wavelet Correlations}

Consider two time series \( \{X_t\} \) and \( \{Y_t\} \), with discrete wavelet transform  coefficients \( \{d_j^X\} \) and \( \{d_j^Y\} \) at decomposition level \( j \). The DWT represents each series as a collection of components associated with distinct frequency bands or scales.

When the DWT is orthogonal, it preserves covariance between the two time series. 
Parseval’s identity guarantees that the scaled sum of the covariances between the corresponding wavelet coefficients over all levels equals the covariance between the original signals:
\begin{eqnarray}
\label{eq:covd} 
\operatorname{Cov}(X_t, Y_t) 
= \frac{1}{2^L}\operatorname{Cov}(c_{J-L}^X, c_{J-L}^Y)
+ \sum_{j=J-L}^{J-1} \frac{1}{2^{J-j}}\operatorname{Cov}(d_j^X, d_j^Y).
\end{eqnarray}
This decomposition follows from the energy-preserving property of orthogonal wavelet transforms, which ensures that the total variance is exactly partitioned among the wavelet coefficients.

These weights depend on the proportion of the total variance at each scale for each series, and in general their sum is not equal to one.

While covariances add cleanly across scales, correlations do not. The correlation at level $j$ is defined as
\begin{eqnarray*}
\rho^*_{J-L} &=& \operatorname{Corr}(c_{J-L}^X, d_{J-L}^Y)
= \frac{\operatorname{Cov}(c_{J-L}^X, d_{J-L}^Y)}{\sigma^*_{X,J-L}\,\sigma^*_{Y,J-L}} \\
\rho_j &=& \operatorname{Corr}(d_j^X, d_j^Y) 
= \frac{\operatorname{Cov}(d_j^X, d_j^Y)}{\sigma_{X,j}\,\sigma_{Y,j}}, 
\quad j = J-L, \dots, J-1; 
\end{eqnarray*}
where $\sigma_{X,j}$, $\sigma_{Y,j}$, $\sigma^*_{X,J-L}$, and $\sigma^*_{Y,J-L}$ are the standard deviations of the wavelet coefficients at the corresponding detail and smooth scales. 
Because these standard deviations vary from one scale to another, the overall correlation between $X_t$ and $Y_t$,
$$
\rho_{X,Y} = \frac{\operatorname{Cov}(X_t, Y_t)}{\sigma_X \sigma_Y},
$$
with $\sigma_X = \sqrt{\operatorname{Var}(X_t)}$ and $\sigma_Y = \sqrt{\operatorname{Var}(Y_t)}$, cannot be recovered by simply averaging the $\rho_j$ values.

Substituting the covariance decomposition (\ref{eq:covd}) into the definition of $\rho_{X,Y}$ gives
$$
\rho_{X,Y}
  = \frac{1}{2^{L}\,\sigma_X \sigma_Y}\operatorname{Cov}(c_{J-L}^X, c_{J-L}^Y)
      + \sum_{j=J-L}^{J-1} \frac{1}{2^{J-j}\,\sigma_X \sigma_Y}\operatorname{Cov}(d_j^X, d_j^Y),
$$
and therefore the correlation between $X$ and $Y$ can be represented as a weighted sum of levelwise correlations,
$$
\rho_{X,Y}
= w^*\,\rho^*_{J-L} + \sum_{j=J-L}^{J-1} w_j\,\rho_j,
$$
where the scale-specific weights are
$$
w_j = \frac{\sigma_{X,j}\,\sigma_{Y,j}}{2^{J-j}\,\sigma_X \sigma_Y},
\qquad
w^* = \frac{\sigma^*_{X,J-L}\,\sigma^*_{Y,J-L}}{2^{L}\,\sigma_X \sigma_Y}.
$$
These weights depend on the proportion of the total variance at each scale for each series and, in general, do not sum to one.


\noindent
{\bf Correlograms.~}To plot levelwise correlograms we need
confidence intervals for levelwise correlations.  
Let $r$ be the sample Pearson correlation from two corresponding wavelet levels of size $n$  and  let $\rho$ the corresponding population parameter. 
The Fisher $z$-transform applied on $r$ 
leads to
$$
w = \tfrac{1}{2}\log\frac{1+r}{1-r} = \operatorname{arctanh}(r).
$$

Under the usual approximation,
$$
w \sim \mathcal{N}\!\left(\operatorname{arctanh}(\rho), \; \tfrac{1}{n-3}\right),
$$
where $\xi$ is the population counterpart of $w$,
$$
\xi = \operatorname{arctanh}(\rho) = \tfrac{1}{2}\log\frac{1+\rho}{1-\rho}.
$$
which leads to the $(1-\alpha) 100\%$ confidence interval for $\xi$,
$$
[w_L, w_U] = \Big[w - \tfrac{z_{1-\alpha/2}}{\sqrt{n-3}}, \; w + \tfrac{z_{1-\alpha/2}}{\sqrt{n-3}}\Big].
$$

Transforming back gives an approximate $(1-\alpha) 100\%$ CI for $\rho$:
\begin{eqnarray}
    \label{cirho}
[r_L, r_U] = [\tanh(w_L), \; \tanh(w_U)] 
= \left[\frac{e^{2w_L}-1}{e^{2w_L}+1}, \; \frac{e^{2w_U}-1}{e^{2w_U}+1}\right].
\end{eqnarray}

When $n$ is small, a bias-corrected version of the transform can improve accuracy:
$$
w_{\text{bc}} = \tfrac{1}{2}\log\frac{1+r}{1-r} - \frac{r}{2(n-1)}.
$$
Use $w_{\text{bc}}$ in place of $w$ in the interval formula above, then transform back with $\tanh(\cdot)$. For more details, see  
\cite{vidakovic2011bioeng}.

\subsection{Partial and Semi-Partial Wavelet Correlations}

Let $X, Y,$  and $Z$ be the sequences of coefficients in  corresponding levels of three wavelet decompositions. The partial correlation between $X$ and $Y$, given  set $Z$, is:

\begin{eqnarray}
\label{partial}
r_{XY.Z} = \frac{r_{XY}-r_{XZ}r_{YZ}}{\sqrt{1-r_{XZ}^2}~\sqrt{1-r_{YZ}^2}}
\end{eqnarray}

The derivation of the $(1-\alpha) 100\%$ confidence interval for partial correlation follows the above arguments leading to the equation in (\ref{cirho}),
but with 
$$
[w_L, w_U] = \Big[w - \tfrac{z_{1-\alpha/2}}{\sqrt{n-p-2}}, \; w + \tfrac{z_{1-\alpha/2}}{\sqrt{n-p-2}}\Big],
$$
where $p$ is the number of variables in 
the control set $Z$. When $Z$ has more than one variable, the relation 
(\ref{partial}) is repeated iteratively. For example, if $Z=(U,V),$
\begin{eqnarray*}
r_{XY.UV} = \frac{r_{XY.U}-r_{XV.U} \cdot r_{YV.U}}{\sqrt{1-r_{XV.U}^2} \cdot \sqrt{1-r_{YV.U}^2}}.
\end{eqnarray*}

\vspace*{0.2in}

 The semipartial correlation controls for $Z$ only in one of the two variables. For instance, the semipartial correlation between $X$ and $Y$ controlling for $Z$ in $X$ is
$$
r_s = r_{XY \cdot Z(X)} = \frac{r_{XY} - r_{XZ} r_{YZ}}
{\sqrt{1-r_{XZ}^2}}.
$$
This quantifies the unique contribution of $X$ to $Y$ after removing the effect of $Z$ from $X$ only. The squared semipartial correlation, $r_s^2$, gives the proportion of variance in $Y$ uniquely explained by $X$ beyond $Z$.

An approximate $(1-\alpha) 100\%$ confidence interval for a semipartial correlation can be obtained via the Fisher $z$-transformation:
$$
w = \tfrac{1}{2} \ln \left( \frac{1+r_s}{1-r_s} \right), \quad
SE(w) \approx \frac{1}{\sqrt{n-p-1}},
$$
where $n$ is the sample size and $p$ is the number of control variables. Then, as in (\ref{cirho}),
$$
[(r_s)_L, (r_s)_R] = \left[
\tanh \left(w - \frac{z_{1-\alpha/2}}{\sqrt{n-p-1}} \right) , \,
\tanh \left(w + \frac{z_{1-\alpha/2}}{\sqrt{n-p-1}}\right)
\right].
$$

\subsection{Beyond Pearson's Correlation}
Another measure of correlation that we apply in the multiscale domain is Kendall's $\tau$. Kendall's $\tau$ is a nonparametric measure of association between two variables, 
based on the relative ordering of paired observations. It measures not only linear association, like Pearson's correlation coefficient, but it is also sensitive to monotonic relations.

For two sequences of size $n$, $X$ and $Y$, consider all pairs $(X_i,Y_i)$ and $(X_j,Y_j)$ with $1 \leq i<j \leq n$:  
 A pair will be called \textbf{concordant} if $(X_i - X_j)(Y_i - Y_j) > 0.$
For \textbf{discordant} pairs it holds $(X_i - X_j)(Y_i - Y_j) < 0$.
We assume no ties since wavelet transforms of numerical sequences have continuous form.

The sample Kendall correlation is
$$
\hat\tau = \frac{C - D}{\binom{n}{2}},
$$
where $C$ and $D$ are the number of concordant and discordant pairs among the total of $\binom{n}{2} = n(n-1)/2$  pairs.  

Kendall's $\tau$ is a robust and distribution-free measure, meaning that it does not require distributional assumptions for $X$ and $Y$.
As we pointed out, it directly measures monotonic association rather than linear correlation.  
 With continuous data, we generally assume no ties, so the formula above applies directly.  
 If ties are present,  corrected versions $\tau_b$ and $\tau_c$ can be used.  

When $c_i$ is the number of concordant pairs involving observation $i$, and let $D$ is the number of discordant pairs, the sample variance of $\hat\tau$ is
$$
\operatorname{Var}(\hat\tau) = 
\frac{4 \sum_i c_i^2 - 2 C - 2 D (2n-3) - C^2 / (n(n-1))}{\binom{n}{2}^2}.
$$
This exact formula accounts for the actual distribution of concordances and discordances in the sample and is especially useful when $n$ is small.

For $n$ large, the sampling distribution of $\hat\tau$ is approximately normal:
$$
\hat\tau \sim N\Big(\tau, \frac{2(2n+5)}{9n(n-1)} \Big).
$$
Based on this asymptotics, the approximate $(1-\alpha)100\%$ confidence interval is
$$
\left[\hat\tau - z_{1-\alpha/2} \sqrt{\frac{2(2n+5)}{9n(n-1)}}, \;\;
 \hat\tau + z_{1-\alpha/2} \sqrt{\frac{2(2n+5)}{9n(n-1)}}\right]
 \cap [-1, 1],
$$
where $z_{1-\alpha/2}$ is the standard normal quantile.  
When the asymptotic variance is replaced with $\operatorname{Var}(\hat\tau)$ we get more accurate confidence interval for small $n$.

\vspace*{0.2in}

\noindent
{\bf Comment.} It is possible to define wavelet correlations using Spearman's measure. The relationship between ranks is linear, data are already continuous without ties, so Pearson correlation directly measures the linear association and is more powerful.
Thus, Spearman’s $\rho$ does not add much extra information in perfectly continuous, linear data, though it is still valid.

Moreover, a broad class of wavelet correlation measures can be expressed within the unifying framework of the so--called \textit{G--correlation}
\parencite{kendall1948rank}. In this formulation, the association between two random variables $X$ and $Y$ is quantified through antisymmetric functions $G_X(\cdot,\cdot)$ and $G_Y(\cdot,\cdot)$, which capture differences or concordance between pairs of observations. The general form is given by
\begin{eqnarray*}
\rho_G(X,Y) = \frac{\mathbb{E}\!\left[G_X(X_1,X_2)\,G_Y(Y_1,Y_2)\right]}{\sqrt{\mathbb{E}\!\left[G_X^2(X_1,X_2)\right]\mathbb{E}\!\left[G_Y^2(Y_1,Y_2)\right]}},
\end{eqnarray*}
where $(X_1,Y_1)$ and $(X_2,Y_2)$ are independent copies of $(X,Y)$. By selecting appropriate $G$--functions, one recovers many classical correlation coefficients as special cases. For instance, choosing $G_X(x_1,x_2)=x_1-x_2$ and $G_Y(y_1,y_2)=y_1-y_2$ yields the \emph{Pearson correlation}, while setting $G_X(x_1,x_2)=\mathrm{sign}(x_1-x_2)$ and $G_Y(y_1,y_2)=\mathrm{sign}(y_1-y_2)$ leads to the \emph{Spearman} or \emph{Kendall} rank correlations depending on normalization. Alternative choices, such as $G_X(x_1,x_2)=F_X(x_1)-F_X(x_2)$ and $G_Y(y_1,y_2)=F_Y(y_1)-F_Y(y_2)$, produce the \emph{Gini correlation}, while $G_X(x_1,x_2)=\mathrm{sign}(x_1-m_X)$ and $G_Y(y_1,y_2)=\mathrm{sign}(y_1-m_Y)$ correspond to the \emph{Blomqvist} median correlation. This unified representation highlights the common structure underlying diverse dependence measures, all interpretable as normalized expectations of products of pairwise contrast functions.

\section{Illustrative Example}

To demonstrate the wavelet correlation methods discussed and highlight their advantage in detecting correlation structures at various frequency scales, we simulate two pairs of a 512-point AR(1) time series according to the following formulas:

$$
\begin{aligned}
X_t & = 0.5X_{t-1} + \epsilon_t\\
Y_t &= 0.5X_{t-1} + 0.5Y_{t-1} + \epsilon_t 
\end{aligned}
$$
and
$$
\begin{aligned}
X_t & = 0.5X_{t-1} + \epsilon_t\\
Y_t &= X_{t-1} + 0.5Y_{t-1} + \epsilon_t 
\end{aligned}
$$

\begin{figure}[h!]
    \centering
    \includegraphics[width=14cm]{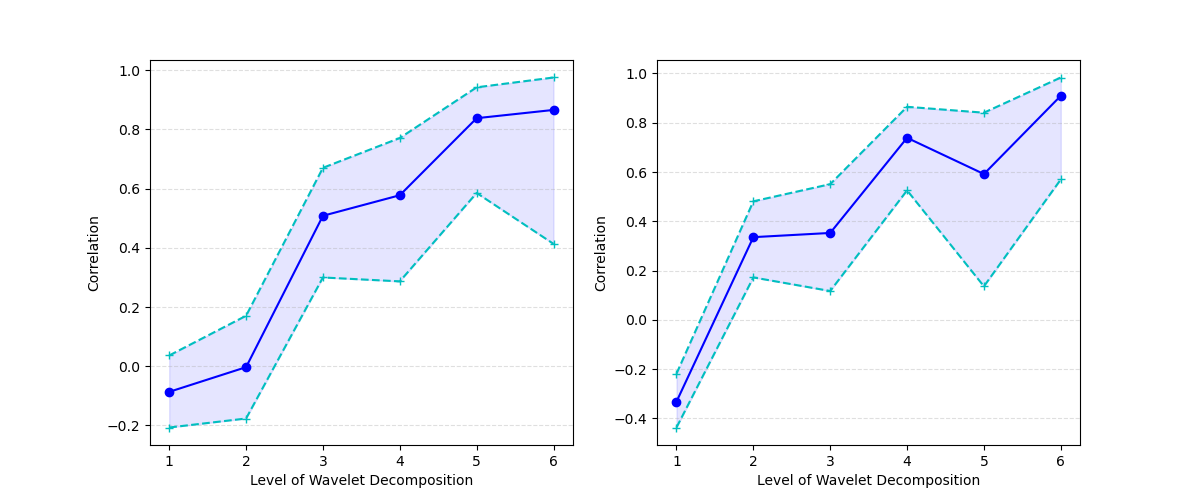}
    \caption{Scale-based, level-wise correlation coefficients for $X_t$ and $Y_t$, systems 1 (Left panel) and 2 (Right panel).}
    \label{fig:AR1_WCC}
    \end{figure}

For each variable, $X_t$ and $Y_t$, a 6-level discrete wavelet decomposition is performed using the Haar wavelet filter. The  correlation is then calculated between the coefficients of the decomposition of $X_t$ and $Y_t$ for each level of  decomposition, along with 95\% confidence interval bands. For comparison, The overall correlation between $X_t$ and $Y_t$ in system 1 is 0.313, and for system 2 it is 0.333. The results are provided in Figure \ref{fig:AR1_WCC}. In this figure, and all figures that follow the scale of the finest level of detail is denoted on the x axis by 1, the next coarser by 2, and so on.

The behavior of the correlation estimates is as expected, and highlights the functionality of the method. We observe that at the finest decomposition scales (levels 1 and 2), the coefficients are largely reflecting the i.i.d. noise inherent in both signals, leading to small and insignificant correlation coefficients. However, because both systems are similar at an average level, and indeed co-dependent, the correlation coefficients jump at the coarser levels of decomposition, reflecting this relationship between the signals. Furthermore, we observe that system 2 displays slightly stronger correlation overall, as well as at the coarsest levels of decomposition, which we expect given the structure of the system. 

This simple illustrative example reflects one of the key advantages of wavelet-based correlation studies, namely, the deconstruction of the overall covariance and correlation into scale-specific components, allowing for a richer scale-dependent understanding of the interrelationships between variables in a complex system.

\section{Applications}
We present two  application examples. The first uses turbulence data collected over a forest clearing, where multiscale correlations between wind components and temperature are analyzed using the Haar wavelet. The second applies two-dimensional NDWT to satellite imagery of the African coast, exploring how correlations between morning, noon, and evening images evolve across scales, and how controlling for one group affects the correlation between others.

In both cases, the wavelet-based approach uncovers scale-dependent relationships that global methods would overlook, providing richer insight into the dynamics of the underlying processes.
\subsection{Turbulence Data Analysis over a Forest Clearing}
Time series measurements of $u$, $w$, and $T$ were collected over a grass-covered forest clearing
at Duke Forest near Durham, North Carolina.  The measurements were collected on June 12-16 at $5.2~m$ above
 the grass surface using a {\sc Gill} triaxial sonic anemometer.  Sonic anemometers measure velocity
by sensing the effect of wind on transit times of sound pulses traveling in opposite directions across
a known instrument distance $d_{sl}$ = ($0.149$ m in this study).  The measurements were sampled at $f_s=56 \mbox{ Hz}$
and were subsequently divided into 19.5 minute intervals to produce $N=65,536$ time measurement
per flow variable per run.  19.5 minute intervals were chosen to ensure stationary conditions within a given run.
 We focus on an ensemble of $103$  runs collected over a wide range
 of $\xi$ ranging from near convective to stable atmospheric flows.   In these runs, the friction velocity ${u_*}$ varied
from $0.04$ ms$^{-1}$ to $0.47$ ms$^{-1}$, and the sensible heat flux varied from $-48$ Wm$^{-2}$ to $369$ Wm$^{-2}$.
In short, the ensemble size exceeds $6.75 \times 10^{6}$ time measurement (but the analysis is conducted on individual
runs prior to ensemble averaging).  Since instrument averaging occurs for separation distances smaller than $d_{sl}$, we restrict the estimation
of $D_q$ to $d_{sl}<r<2.5 m$.  The time series was converted to one-dimensional cuts through the flow by using the frozen
turbulence hypothesis \parencite {PraMenSre1988}.  That is, $dx = \overline U f_s^{-1}$, where $\overline U$ is the mean wind speed.
Further details about the experimental setup, atmospheric conditions, inertial subrange identification,
and instrumentation details can be found elsewhere \parencite{KatulHsiehSigmon1997, KatulAlbertsonVidakovic2000}. 

\begin{figure}[h!]
    \centering
    \includegraphics[width=10.5cm]{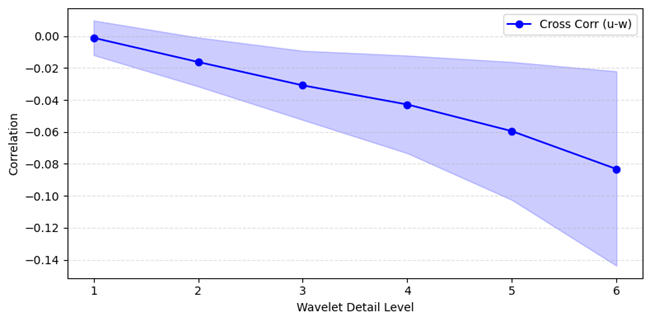}
    \caption{95\% C.I for cross correlation between $u$ and  $w.$}
    \label{fig:ci_cross}
    \end{figure}

\begin{figure}[h!]
    \centering
    \includegraphics[width=10.5cm]{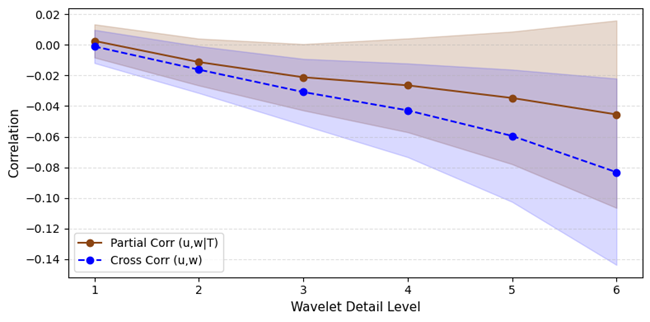}
    \caption{95 \% confidence intervals for the partial cross-correlation between $u$ and $w,$ given $T.$
    The solid dark-brown line represents the mean partial correlation across wavelet detail levels 1-6, and the dashed blue line shows the corresponding cross-correlations from Figure \ref{fig:ci_cross}.}
    \label{fig:cross_pcross_ci}
    \end{figure}

\begin{figure}[h!]
    \centering
    \includegraphics[width=10.5cm]{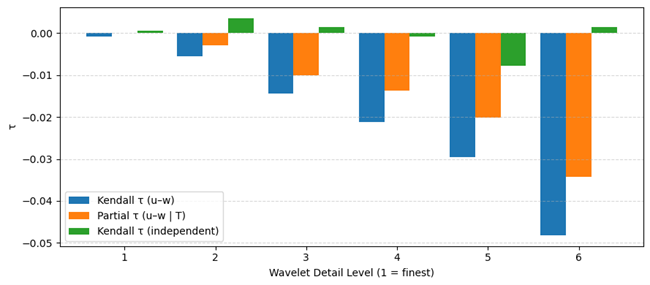}
    \caption{Average Kendall tau correlation between $u$ and $w$ (same run), partial Kendall tau correlations given $T$, and average Kendall tau correlation between $u$ and $w$ (from different runs). }
    \label{fig:kendall_corr}
\end{figure}

\begin{figure}[h!]
    \centering
    \includegraphics[height=5cm,width=10.5cm]{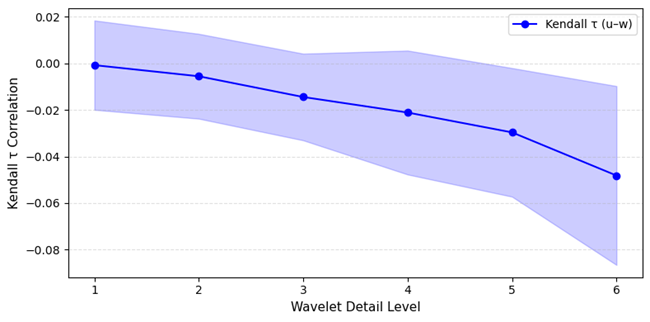}
    \caption{95\% C.I for Kendall tau correlation between $u$ and $w$.}
    \label{fig:kendall_ci}
    \end{figure}

We analyzed ten atmospheric turbulence records, each containing simultaneous measurements of the three velocity components $u$, $v$, $w$ and the temperature $T$ (in Kelvin). We used the 65,536 samples from each series to capture the full multiscale structure. We then performed a six-level discrete Haar wavelet decomposition on the $u$, $w$, and $T$ signals. At each wavelet scale, we computed both the raw Pearson correlation coefficient corr($u$, $w$) and the partial correlation corr($u$, $w$ $|$ $T$). \\
   Then we took the $u$ components from the first ten files and the $w$ components from a separate set of ten files, and computed the cross‑correlation between these independent $u$ and $w$ series (Table \ref{tab:average_cross}). 
   \begin{table}[h]
   \centering
\begin{tabular}{|c|c|c|c|}
\hline
\textbf{Level} & \textbf{\begin{tabular}[c]{@{}c@{}}Average cross\\  correlation\end{tabular}} & \textbf{\begin{tabular}[c]{@{}c@{}}Average partial\\  cross correlation\end{tabular}} & \textbf{\begin{tabular}[c]{@{}c@{}}Average cross\\  correlation (independent)\end{tabular}} \\ \hline
1              & -0.001120                                                                     & 0.002496                                                                              & -0.000096                                                                                   \\ \hline
2              & -0.016236                                                                     & -0.011301                                                                             & 0.000164                                                                                    \\ \hline
3              & -0.030862                                                                     & -0.021217                                                                             & -0.000724                                                                                   \\ \hline
4              & -0.042834                                                                     & -0.026520                                                                             & 0.000195                                                                                    \\ \hline
5              & -0.059505                                                                     & -0.034745                                                                             & -0.000633                                                                                   \\ \hline
6              & -0.083216                                                                     & -0.045505                                                                             & -0.001460                                                                                   \\ \hline
\end{tabular}
\caption{Average cross correlation between $u$ and $w$ (same run), partial correlations given $T$, and average cross correlation between $u$ and $w$ (from different runs) }
\label{tab:average_cross}
\end{table}\\
   The following Figure \ref{fig:cross_cor} shows, for wavelet levels 1-6, the average cross correlation between 
$u$ and $w$ from the same record (blue), the partial correlation controlling for $T$ (orange), and the cross‑correlation between $u$ and $w$ from independent datasets (green).
\begin{figure}[h!]
    \centering
    \includegraphics[width=10.5cm]{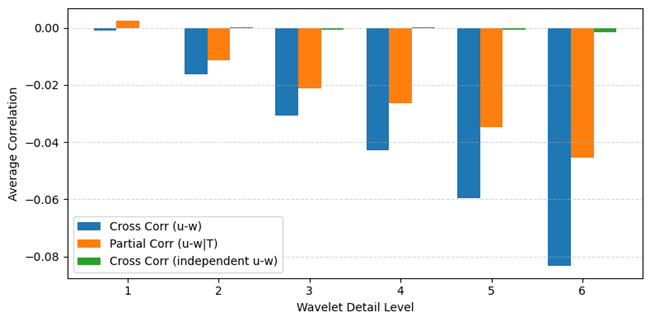}
    \caption{Average cross correlation between $u$ and $w$ (same run), partial correlations given $T$, and average cross correlation between $u$ and $w$ (from different runs). }
    \label{fig:cross_cor}
    \end{figure}\\
Figure \ref{fig:ci_cross} shows the average cross correlation between the wavelet detail coefficients of $u$ and $w$ from the same dataset, plotted at levels 1 (finest) through 6 (coarsest). The solid line marks the mean correlation over ten records, and the shaded band is the 95\% confidence interval obtained via Fisher’s $z$ transform.\\
\\ 
Figure \ref{fig:cross_pcross_ci} compares the cross and partial correlations between $u$ and $w$ across wavelet detail levels 1-6. Both measures decrease with increasing wavelet detail level, indicating stronger anti-correlation at coarser scales. The partial correlations (solid dark-brown line) are consistently less negative than the cross-correlations (dashed blue line), suggesting that temperature explains part of the association between the two components. 
 In the surface layer of the atmospheric boundary layer, the horizontal velocity component $u$ and the vertical velocity component $w$ are often negatively correlated. Rising air parcels ($w$) tend to originate near the surface and generally have lower horizontal speeds ($u < \overline{u}$), while descending parcels ($w < 0$) tend to come from higher altitudes where the mean wind speed is greater ($u > \overline{u} $). This leads to a negative covariance \( \overline{u'w'} < 0 \), which corresponds to a downward turbulent flux of horizontal momentum, expressed as the Reynolds stress component $\tau = -\rho \, \overline{u'w'}$.\\
By contrast, the vertical velocity and temperature ($w$ and $T$) are generally positively correlated in convective daytime conditions: warmer air parcels near the heated surface rise ($w > 0, T > \overline{T}$) and cooler air aloft sinks ($w < 0,  T < \overline{T}$). This yields $ \overline{w' T'} > 0 $ and a heat flux $Q_H = \rho c_p \, \overline{w'T'}$. Where \( \rho \) is air density and \( c_p \) is the specific heat at constant pressure. A positive \( Q_H \) indicates net upward heat transfer, characteristic of daytime surface heating. These covariances negative \( \overline{u'w'} \) and positive \( \overline{w'T'} \) are hallmarks of turbulent transport in the atmospheric boundary layer and are essential in modeling momentum and heat exchange between the surface and the atmosphere. \\ \\
Then, at every detail level $j=1,\dots, 6$, we quantified the monotonic association between $u$ and $w$ using Kendall’s tau computed on the corresponding detail coefficients. The following Table \ref{tab:kendalltau_corr} and Figure \ref{fig:kendall_corr} show the average Kendall tau correlation between $u$, $w$ (same sample), average Kendall tau partial correlation, and average Kendall tau correlation between $u$, $w$ for independent data:  
\begin{table}[h]
\centering
\begin{tabular}{|c|c|c|c|}
\hline
\textbf{Level} & \textbf{\begin{tabular}[c]{@{}c@{}}Average Kendall \\ tau (u-w)\end{tabular}} & \textbf{\begin{tabular}[c]{@{}c@{}}Average  Kendall tau \\Partial correlation\end{tabular}} & \textbf{\begin{tabular}[c]{@{}c@{}}Average Kendall \\ tau (independent u-w)\end{tabular}} \\ \hline
1              & -0.000764                                                                     & 0.000040                                                                        & 0.000602                                                                                  \\ \hline
2              & -0.005540                                                                     & -0.002889                                                                       & 0.003542                                                                                  \\ \hline
3              & -0.014449                                                                     & -0.009988                                                                       & 0.001411                                                                                  \\ \hline
4              & -0.021121                                                                     & -0.013798                                                                       & -0.000876                                                                                 \\ \hline
5              & -0.029641                                                                     & -0.020143                                                                       & -0.007750                                                                                 \\ \hline
6              & -0.048168                                                                     & -0.034240                                                                       & 0.001385                                                                                  \\ \hline
\end{tabular}
\caption{Average Kendall tau correlation between $u$ and $w$ (same run), partial Kendall tau correlations given $T$, and average Kendall tau correlation between $u$ and $w$ (from different runs). }
\label{tab:kendalltau_corr}
\end{table}
Figure \ref{fig:kendall_ci} shows the average Kendall tau correlation between the wavelet detail coefficients of $u$ and $w$ from the same dataset, plotted at levels 1 (finest) through 6 (coarsest). The solid line marks the mean Kendall tau correlation over ten records, and the shaded band is the 95\% confidence interval obtained via Fisher’s $z$ transform.
\\
\subsection{Two-dimensional NDWT Analysis of African Coast Satellite Imagery} 
\begin{figure}[h!]
    \centering
    \includegraphics[width=12cm]{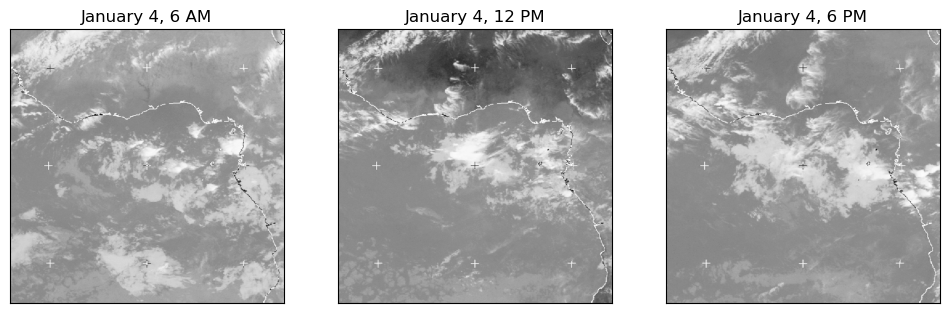}
    \caption{A typical observation: images  from morning, noon, and evening groups. }
    \label{fig:sampleimg}
\end{figure}

For this example, we focus on cross-correlation and partial cross-correlation of wavelet coefficients from a 2-dimensional discrete wavelet transform. Our data consists of 36 infrared satellite images of the west coast of Africa and the Gulf of Guinea, taken on 12 consecutive days (1/4/2001- 1/15/2001). The brightness of pixels in the image represents the intensity of infrared radiation reflected by the Earth's surface or the tops of clouds, which corresponds to its temperature. The images are cropped to squares of size $512 \times 512$. Each day, 3 images were taken, at 6:00 AM, 12:00 PM, and 6:00 PM. These images are referred to as the Morning, Noon, and Evening groups. A typical example of a single day's images is demonstrated in Figure \ref{fig:sampleimg}.

Of interest is the following:

(i) Can obtaining the correlation coefficients between each day's images at each level of the wavelet decomposition provide further information about structures and behaviors in the images present at different scales, which might be overlooked with a simple global correlation coefficient?

(ii) Does controlling for one group (say, morning) when calculating correlation between the other groups (noon and evening) provide more information about these correlation behaviors at various scales?
 
\vspace*{0.15in}

Here, a 5-level wavelet decomposition is performed using the Coiflet wavelet filter, commonly used for 2D wavelet decomposition and chosen for its symmetry. Unlike the previous example, a non-decimated or stationary wavelet transform is used. For each day's images, the 5-level NDWT is performed and coefficients are obtained. From these coefficients, we calculate both the Pearson correlation between each group, ($corr(u, v)$), and the partial correlation between groups, ($corr(u, v | w)$). Then, each of these correlation coefficients are averaged over the 12-day sample to provide an overall mean correlation estimate and its 95\% confidence interval. 

\begin{figure}[h!]
    \centering
    \includegraphics[width=14cm]{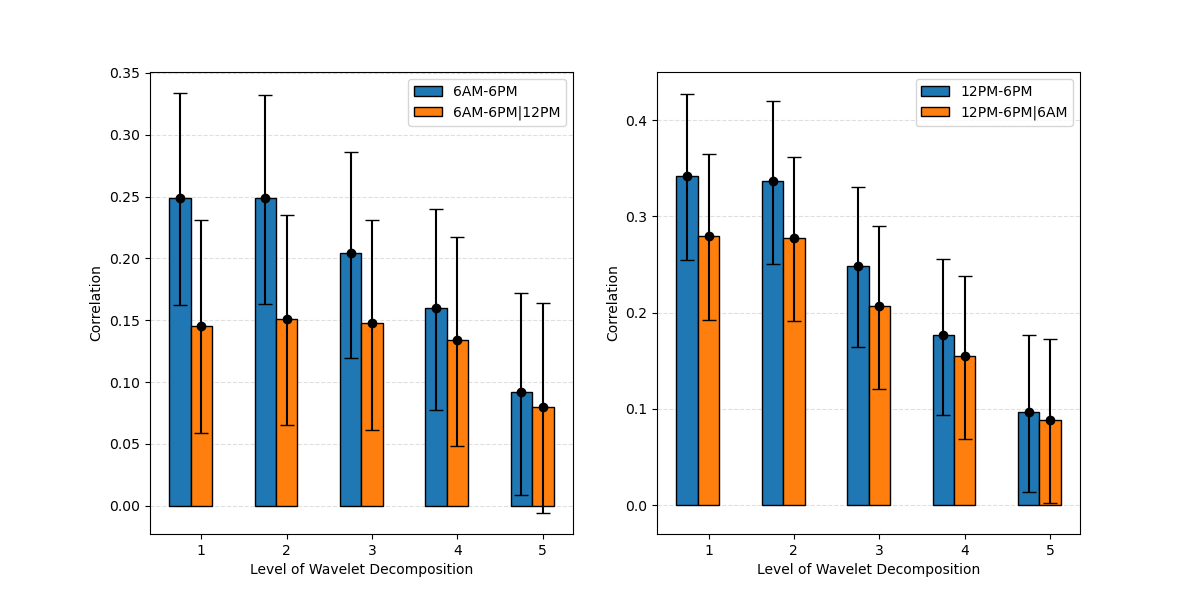}
    \caption{Average 2D wavelet correlation and partial correlation, with 95\% confidence intervals. Left panel: morning-evening and morning-evening given noon;  Right panel: noon-evening and noon-evening given morning.}
    \label{fig:corr_pcorr_CI}
\end{figure}

Figure \ref{fig:corr_pcorr_CI} shows the obtained coefficient estimates and their confidence intervals for two image groups: the correlation between the 6 AM images and 6 PM images, before and after controlling for the influence of the 12 PM images, and the 12 PM and 6 PM images, before and after controlling for the influence of the 6 AM images.

The results in Figure \ref{fig:corr_pcorr_CI} demonstrate that at finer scales of the images, corresponding to the lower levels of the decomposition, controlling for the third group has a pronounced effect on the correlation between images from the other groups, with this effect diminishing as the scale increases. Unsurprisingly, the noon and evening groups show stronger correlation at each level, even after controlling for the influence of the morning group, with the morning group having a weaker effect on the other two. We also observe that at coarser scales (levels 3-5), the level of correlation decreases, indicating that much of the correlation between the image groups occurs due to similar small-scale behaviors in the images, with less similarity as the scale of analysis increases. 

\section{Discussion}

Wavelet correlations provide a scale-resolved view of dependence that a single global statistic cannot. By decomposing signals into orthogonal components, the DWT partitions covariance exactly across scales, revealing where two processes co-vary most strongly. In practice, this often shows that global correlations are driven by only a few scales, or conversely that weak overall association masks concentrated scale-specific dependence. Such localization is valuable in settings where mechanisms act differently across frequencies, such as turbulence, finance, microeconomics, geoscience, and medical signal or image processing.

A key advantage of the orthogonal transform is interpretability: sums of cross-products are preserved and decomposed additively across scales. Correlations themselves are not additive, but when adjusted for sample sizes, the overall correlation is recovered as a weighted sum of levelwise correlations, making the contribution of each scale explicit.

Partial and semipartial correlations extend naturally to the wavelet domain and are practically useful. They reveal conditional associations confined to certain scales and separate direct from mediated effects, offering insight unavailable from global measures.

Orthogonal and non-decimated transforms each have trade-offs, energy partitioning versus translation invariance, and the choice should match the scientific question. Beyond Pearson correlation, rank-based and kernel dependence measures can also be embedded in the wavelet framework, combining robustness or generality with scale localization. Applications extend across disciplines, from turbulence and climate science to finance, economics, neuroscience, and medicine, where the aim is not just whether signals relate, but at which scales.

In sum, wavelet correlations turn a single global number into a structured spectrum of correlations, highlighting mechanisms and mediators that would otherwise remain hidden. The task for practitioners is to choose the transform suited to their problem, report scale contributions transparently, and interpret wavelet-domain measures as complements to, not replacements for, time-domain analysis.

In the interest of reproducibility, all code developed for this analysis is available in the form of Jupyter notebooks at {\tt https://github.com/jgkissell/WaveletCrossCorrelations}.

\paragraph{Acknowledgments.}
B. Vidakovic acknowledges the partial support of the H.O. Hartley Chair foundation and NSF Award 2515246 at Texas A\&M University.


 
\printbibliography

\end{document}